%% file: main.tex
\documentclass[a4paper]{article}

\usepackage{INTERSPEECH2019}
\usepackage{times}  
\usepackage{helvet}  
\usepackage{courier}  
\usepackage{url}  
\usepackage{graphicx}  
\usepackage{todonotes}
\usepackage[english]{babel}
\usepackage[utf8]{inputenc}
\usepackage{algorithm}
\usepackage[noend]{algpseudocode}
\usepackage{amssymb}
\usepackage{array}
\usepackage{multirow}
\usepackage[strings]{underscore}
\graphicspath{{./images/}}
\usepackage{amsmath}
\usepackage{comment}

\usepackage{dblfloatfix}

\title{Adversarial Black-Box Attacks on Automatic Speech Recognition Systems using Multi-Objective Evolutionary Optimization}
\name{Shreya Khare$^1$,
Rahul Aralikatte$^2$,
Senthil Mani$^1$}

\address{
  $^1$IBM Research, $^2$University of Copenhagen}
  \email{skhare34@in.ibm.com,
rahul@di.ku.dk, 
sentmani@in.ibm.com}

\begin{document}

\maketitle

\begin{abstract}

Fooling deep neural networks with adversarial input have exposed a significant vulnerability in the current state-of-the-art systems in multiple domains. Both black-box and white-box approaches have been used to either replicate the model itself or to craft examples which cause the model to fail. In this work, we propose a framework which uses multi-objective evolutionary optimization to perform both targeted and un-targeted black-box attacks on Automatic Speech Recognition (ASR) systems. 
We apply this framework on two ASR systems: Deepspeech and Kaldi-ASR, which increases the Word Error Rates (WER) of these systems by upto 980\%, indicating the potency of our approach. During both un-targeted and targeted attacks, the adversarial samples maintain a high acoustic similarity of $0.98$ and $0.97$ with the original audio.
\end{abstract}

\input{introduction}

\input{approach-new}
\input{experiments}

\input{conclusion}

\bibliographystyle{IEEEtran}
\bibliography{ref}
\end{document}

%% file: introduction.tex
\section{Introduction}
Advancements in deep learning have improved the state-of-the-art systems in domains like computer vision, natural language processing and speech recognition. Recent studies \cite{eva_attack,carlini_NN,Szegedy} have shown that these systems can be easily fooled with carefully crafted inputs. For example, \cite{Goodfellow,gradient_adv} showed that an image classifier could be fooled into classifying an image to a label of choice by introducing small perturbations to the input image which are imperceptible to humans. 
ASR systems are becoming ubiquitous with the pervasiveness of smart devices. With the success of digital assistants, voice is replacing text as the main mode of communication with these devices. Along with automating mundane human activities, these digital assistants can now accomplish complex tasks which may require sensitive user data. For example, Alexa \cite{alexa}
can now access a user's bank details and perform actions like checking the balance and making credit-card payments. Therefore, it is essential that ASR systems are not susceptible to adversarial attacks. Attacks on ASR systems can broadly be classified based on two attributes: i) model transparency, and ii) intent of the attack.

\textit{Attacks based on model transparency}: Attacks in which the adversary has access to the internals of an ASR system (network structure, weights, etc.) are known as \textit{White Box} attacks \cite{carlini_NN}.
Attacks of this kind are rare as commercial ASR systems seldom expose their internal working to users. On the other hand, \textit{Black Box} attacks, where the attacker only has access to the input-output pairs of the model, are more probable. 
\begin{figure}[h]
    \centering
    \includegraphics[width=0.75\columnwidth]{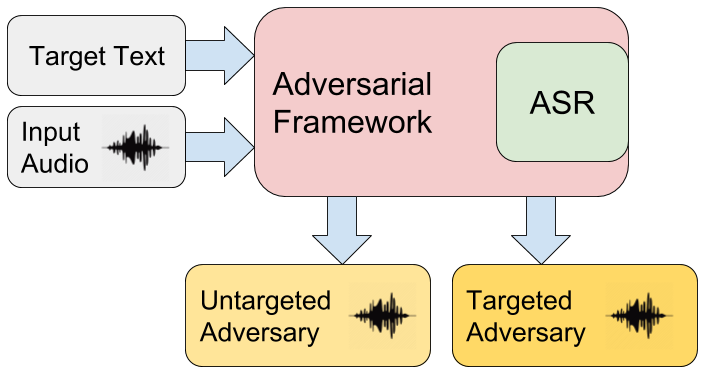}
    \caption{Overview of the adversarial example generation framework for ASR systems}
    \label{fig:overview}
\end{figure}
\textit{Attacks based on intent}: The intention of an adversarial attack can be either to cause failure or manipulate the system towards an end-goal. The former is known as \textit{un-targeted} attacks where the intent is to make the ASR system generate wrong output for a given audio sample. In the latter, known as \textit{targeted} attacks, the aim is to perturb the input in such a way that the ASR outputs a desired text. While un-targeted attacks just degrade an ASR system's performance, targeted attacks are dangerous as they can manipulate actions taken by digital assistants.


Previous works like \cite{Iter2017GeneratingAE} generate adversarial samples for ASR systems by adding perturbations to MFCC features and perform lossy reconstruction of the speech signal in the time domain. This lossy reconstruction makes the perturbations perceptible to the human ear. 
~\cite{carlini}  proposes a method which enables the propagation of gradients to the MFCC reconstruction layer. They perform white box targeted attacks on Mozilla's Deepspeech
model \cite{deepspeech} and obtain adversarial samples which are $99.9\%$ similar to the desired target text. Furthermore, \cite{backdoor,dolphinattack} utilise non-linearities of microphones in smart devices to generate spurious commands (adversarial examples with frequency $~40$ Khz.) and issue malicious commands to digital assistants. Another kind of attack proposed in ~\cite{hiddenvoicecommands}, intends to generate sounds that are interpreted as voice commands by devices but are unrecognisable to humans. These attacks, although powerful, are usually easy to detect and can be prevented using filters.  
\cite{Alzantot_bb} use a genetic algorithm to perform black-box attacks, with an intent of mis-classifying audio samples containing short speech commands.  \cite{Rohan_BB} extended this approach to work on longer phrases and sentences. Their approach is limited to ASR systems which give access to the last layer (logits) and requires the knowledge of the model's loss function. Hence it can be considered as a \textit{grey-box} attack at best. 

In this work, we consider adversarial audio generation as a multi-objective optimization problem which achieves a balance between conflicting objectives: (i) increasing text dissimilarity, and (ii) maintaining acoustic similarity. We propose a framework which is algorithm agnostic and generates near-optimal solutions in both un-targeted and targeted settings. We show that our approach works well with two genetic optimizers: (i) a simple Multi-Objective Genetic Algorithm (MOGA)~\cite{MOGA}, and (ii) Elitist Non-dominated Sorting Genetic Algorithm (NSGA-II)~\cite{NSGA2}. We perform black-box attacks on two popular ASR systems:Deepspeech \cite{deepspeech} and Kaldi-ASR (TDNN chain model) \cite{kaldiasr}, and present both empirical and human evaluation results \footnote{The project web page is at https://shreyakhare.github.io/audio-adversarial/}




\begin{figure*}[hbpt]
    \centering
    \includegraphics[scale=0.4]{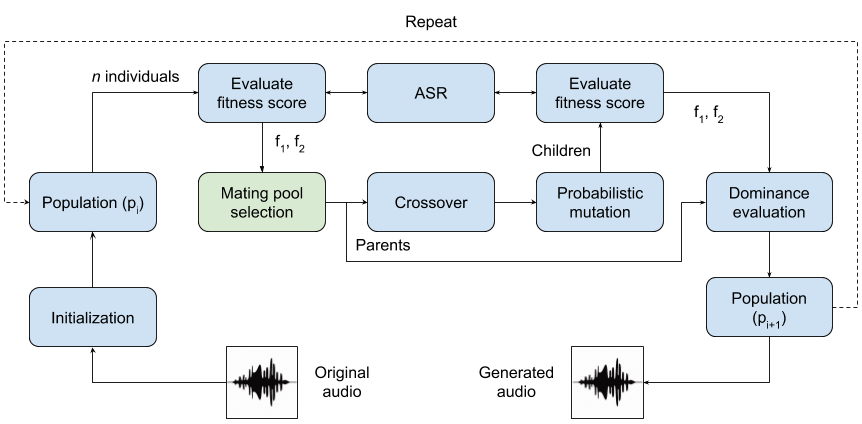}
    \caption{System Architecture}
    \label{fig:system}
\end{figure*}



%% file: approach-new.tex
\section{Approach}
\label{sec:approach}

Multi-objective optimization is a process of finding solutions which must satisfy more than one conflicting objectives. In such cases, often there is no single optimal solution. The interaction among different objectives gives rise to a set of compromised solutions known as trade-off, non-dominated, non-inferior or \textit{Pareto-optimal} solutions \cite{pareto}.\footnote{We use the terms `Solution' and `Individual' interchangeably}

\begin{figure}
    \centering
    \includegraphics[scale=0.35]{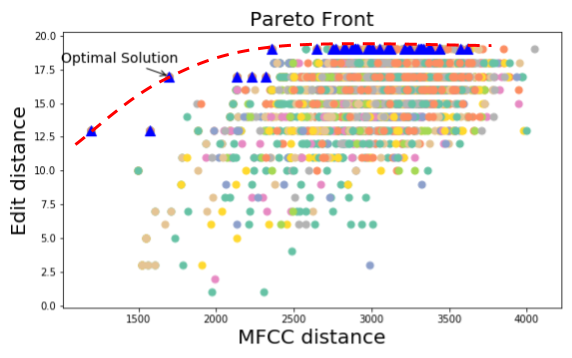}
    \caption{A pareto front showing the dominance of individuals  for  a  random  audio  sample}
    \label{fig:pareto_plot}
\end{figure}

Before moving forward, we define the following terms: (i) \textit{dominance}: $a$ is said to dominate $b$ iff $a$ is no worse than $b$ in every fitness objective (blue points in Fig. \ref{fig:pareto_plot}). If this constraint does not hold between $a$ and $b$, then they are called non-dominating individuals, (ii) \textit{pareto front}: a set of mutually non-dominating individuals all of which are pareto-optimal (individuals on the red line in Fig. \ref{fig:pareto_plot}). There can be multiple pareto fronts in the search space, and (iii) \textit{crowding distance}: an individual is said to have high crowding distance if there are many individuals in its vicinity. It indicates the density of the area around the individual in the search space. For more details, the reader is advised to refer~\cite{CoelloCoello1999}.

As with any evolutionary system, in each generation our framework scores candidate solutions to determine which individuals are the fittest. They are allowed to reproduce using crossover and mutation which results in fitter individuals (better solutions) over time. The approach to generate adversarial samples is shown in Fig.~\ref{fig:system}. The blue boxes represent algorithm agnostic parts of the framework. Only the \textit{mating pool selection} changes with the choice of algorithm. 


\input{exampletable.tex}

\textbf{Evaluation of fitness score} is arguably the most important step in evolutionary algorithms. It decides which individuals survive and therefore prunes the search space. Our goal is to generate audibly similar samples which maximize errors in ASR systems. Parameterizing a fitness function to evaluate this criterion is hard. Therefore we breakdown the fitness function into two components, each of which quantifies a desired property: (i) acoustic similarity -- the Euclidean distance between MFCC \cite{mfcc} of the original and generated audio samples, and (ii) text dissimilarity -- \textit{edit distance} between the texts generated by the ASR system when the original and generated samples are provided as inputs. To the best of our knowledge, we are the first to use such a multi-objective fitness function to optimize an adversarial audio generator.

In the un-targeted setting, the generated sample is considered fitter if it has high acoustic similarity with the input audio and if the text generated has high edit distance with respect to the ASR output. Whereas in the targeted setting, a fit solution, while having high acoustic similarity with the input audio, should also have low edit distance when compared with the desired text.


Once fitness values are computed for every candidate solution, we use an ensemble of techniques for \textbf{mating pool selection}. For MOGA, we define three selection schemes: (i) \textit{Same Rank Selection}: In this selection scheme, we first create two lists $l_1$ and $l_2$ by ranking the individuals based on each of the two fitness criteria mentioned above. Let $l_x \gets i_{xr}$ where $i$ is an individual in list $x$ with rank $r$. We pair two $`i'$s which have the same $r$ in the two lists, (ii) \textit{Inverse Rank Selection}: This scheme is the opposite of the previous one. One list is ranked from best to worst and the other vice-versa. This makes sure that diversity is maintained in the population and we do not constrain the search space too soon, and (iii) \textit{Roulette Wheel Selection} \cite{roulettewheel}: In this scheme, the fitness score is used to compute a probability of selection for each individual. If $f_{i}$ is an individual $i$'s fitness score, then its probability of being selected is: 
\begin{equation}
    p_{i}=\frac{f_{i}}{\sum_{i=1}^N f_{i}}
\end{equation}
where $N$ is the number of individuals in the population. While unlikely, individuals with high fitness scores may still be eliminated ensuring there is no premature convergence to local optima.

Now, the NSGA-II selection procedure can be described as follows: (i) select the $k$ best individuals and move them to the next generation without change, (ii) sort the other individuals by their dominance and pick $N-k$ individuals, and (iii) if candidates of one pareto front are  dominating, use crowding distance to re-sort the individuals and maintain diversity 

Once the mating pairs are identified using the above selection strategies, we use arithmetic recombination operators to perform a single-point \textbf{crossover} \cite{gen_algo}. If $(p_1, p_2)$ is a mating pair, then three children (C) are created where one has the characteristics of the parents in equal proportion, and two are dominated by each parent. 
\begin{equation}
   C=\frac{p_{1}+p_{2}}{2} \bigcup 
   \frac{2 \times p_{1}+p_{2}}{3} \bigcup
   \frac{ p_{1}+2 \times p_{2}}{3}
\end{equation}

Thereafter, we perform \textbf{mutation} on each parameter of every individual by adding Gaussian noise with a probability $prob_m$ so that the new population is scattered over a large area in a different part of the search space thus reducing crowding distance. Once a set of offspring is created, their fitness scores are evaluated in the same way as described previously. The combined set of parents and children are ranked using (i) dominance scores in MOGA, (ii) dominance and crowding distance in NSGA-II . The top $N$ individuals are carried forward to the next generation. This process is repeated until we reach termination.

%% file: exampletable.tex
\begin{table*}[!b]
\centering
\begin{tabular}{|c|c|c|c|c|}
\hline
                              \textbf{Attack} & \textbf{Original text}        & \begin{tabular}[c]{@{}c@{}}\textbf{Generated Text} \\ \textbf{(original)} \end{tabular}    & \textbf{Target Text}  & \begin{tabular}[c]{@{}c@{}}\textbf{Generated Text} \\ \textbf{(attacked)} \end{tabular} \\ \hline
\textit{Un-targeted} & the one you are blocking    & the one you are blocking    & N.A.            & the money of locking    \\ \hline
\textit{Un-targeted} & never mind about that & never mind about that & N.A.            & no i about that    \\ \hline
\textit{Un-targeted} & what's your name he asked   & whats yo name he asked      & N.A.            & one of late ask         \\ \hline \hline
\textit{Targeted}    & i've got to go to him       & ive got to go to him        & a cat           & a cat                   \\ \hline
\textit{Targeted}    & follow the instruction here & follow the instruction here & all of these            & all of these  shapes   \\ \hline
\textit{Targeted}    & do you know what            & did you know that           & that I love you & I love                  \\ \hline
\end{tabular}
\caption{Texts generated by ASR systems on randomly selected audio samples  and their adversarial counterparts}
\label{table:samples}
\end{table*}

%% file: experiments.tex
\section{Experiments and Results}

We use the standard Mozilla Common Voice dataset (CVS) \cite{cvsdataset} to test the effectiveness of our approach. Following \cite{carlini,Rohan_BB}, we take $100$ instances from CVS and use them to perform both un-targeted (make the ASR output as erroneous as possible) and targeted (make the ASR generate a desired output) attacks. Table \ref{table:samples} shows some examples of the original text generated by the ASR systems (on the original audio sample), the target text (the end goal) and the text generated from the adversarial input.  In this section, we present our implementation details along with evaluation methods. We evaluate our approach in two ways: i) \textit{quantitative} -- where we empirically prove the effectiveness of the samples using WER and correlation coefficient (CC), and  ii) \textit{qualitative} -- where we show that humans find it difficult to distinguish between the original and adversarial samples by conducting a survey.


\subsection{Implementation Details}
For each audio sample from the dataset, a set of $100$ individuals are created to form the initial population. They are initialized by adding random uniform noise to the original audio signal with a sampling rate of $16,000$. Therefore, a two-second audio clip will result in individuals having $32,000$ genes. Intuitively, this helps in faster convergence when compared to random initialization. 

After computing the fitness scores based on acoustic (MFCC) and textual (edit distance) similarities, the population is converted to a list of mating pairs. The extraction of MFCC features is performed with window size of $25ms$ and a stride of $10ms$. It is to be noted that an individual may appear in more than one mating pair. The three crossover operations described previously are applied to each pair to produce three children. Probabilistic mutation is applied to every gene of every individual with $prob_m = 0.005$.

Next, the parents and children are combined and re-ranked based on their dominance (and crowding distance in case of NSGA-II) and the top $30$ pareto-optimal individuals are propagated to the next generation. This process is repeated for a $maxiters = 50$, or until we get the same set of pareto-optimal solutions over two successive generations which indicates convergence. The most dominant individual from the last generation is the adversarial counterpart for our input audio signal. After termination, there may be multiple, equally good dominant solutions to pick from (with equal dominance). In this case, we choose one at random as the best solution.

\subsection{Quantitative Evaluation}

\begin{table}[]
\centering
\begin{tabular}{|c|c|c|c|c|}
\hline
\begin{tabular}[c]{@{}l@{}} Attack\end{tabular}                     & Metric               & ASR        & \begin{tabular}[c]{@{}c@{}}MOS+\\MOGA \end{tabular} & \begin{tabular}[c]{@{}c@{}}MOS+\\NSGA\end{tabular} \\ \hline
\multirow{4}{*}{\begin{tabular}[c]{@{}l@{}}Untargeted\\ \end{tabular}} & \multirow{2}{*}{WER $\uparrow$} & Kaldi-ASR  & 3.9                                                            & \textbf{4.68}                                                \\ \cline{3-5}           &      
& Deepspeech & \textbf{5.4}                                                            & 3.7                                                 \\ \cline{2-5} & \multirow{2}{*}{CC $\uparrow$ }  & Kaldi-ASR  & \textbf{0.97}                                                           & \textbf{0.97}                                                \\ \cline{3-5} &                      & Deepspeech & \textbf{0.98}                                                           & \textbf{0.98}                                                \\ \hline
\multirow{4}{*}{\begin{tabular}[c]{@{}l@{}}Targeted \\ \end{tabular}}  & \multirow{2}{*}{WER  $\downarrow$ } & Kaldi-ASR  & \textbf{3.33}                                                           & \textbf{3.33}                                                \\ \cline{3-5}  &  
& Deepspeech & 2.14                                                           & \textbf{1.5}                                                 \\ \cline{2-5} 
& \multirow{2}{*}{CC $\uparrow$}  & Kaldi-ASR  & \textbf{0.98}                                                           & \textbf{0.97}                                                \\ \cline{3-5}   &                      & Deepspeech & \textbf{0.98}                                                           & \textbf{0.97}                                                \\ \hline
\end{tabular}
\caption{Mean word error rates and correlation coefficients of the ASR systems on original and adversarial samples. MOS stands for multi-objective selection. ($\uparrow$ - higher the better, $\downarrow$ - lower the better).}
\label{table:WER}
\end{table}

For quantitative evaluation, we first compute the WER of the original samples using an ASR System. For the selected $100$ samples the WER of Deepspeech is $0.5$ and Kaldi-ASR is $1.0$. After both un-targeted and targeted attacks are performed, the new WER and CC are obtained for the adversarial samples which can be seen in Table \ref{table:WER}. For Deepspeech (Kaldi-ASR), the WER degrades by 980\% (368\%) during un-targeted attacks.

For targeted attacks, the target text snippets are generated as follows: if the word length of the original sample text is $n$, then a phrase whose length is the range of $[2, n+1]$ is chosen randomly from the data to form the target text. In this situation, the $WER$ between the original and targeted text will be in the range $(n, n+1) \approx (5, 6)$. In this setting, the text generated by Deepspeech (Kaldi-ASR) for the adversarial audio samples, are on average $1.0$ ($1.83$) words away from their target texts. This suggests that even after only $50$ generations, the generated samples are quite close to their target. Table \ref{table:WER} also shows that average CC is extremely high for both ASR systems, thus indicating acoustic similarity between the original and adversarial samples.

To compare our results with a previous state-of-the-art, we use the model proposed in \cite{Rohan_BB}. This \textit{grey-box} system (it requires access to the model's loss function and final layer logits) can only perform targeted attacks. Using the same $100$ audio samples and target texts, the mean WER of this system is observed to be $5.0$ and the CC is $0.99$. Compared to this, our framework generates adversarial samples which are much closer to their targets while having comparable acoustic similarity.

\begin{figure}[]
    \includegraphics[scale=0.45]{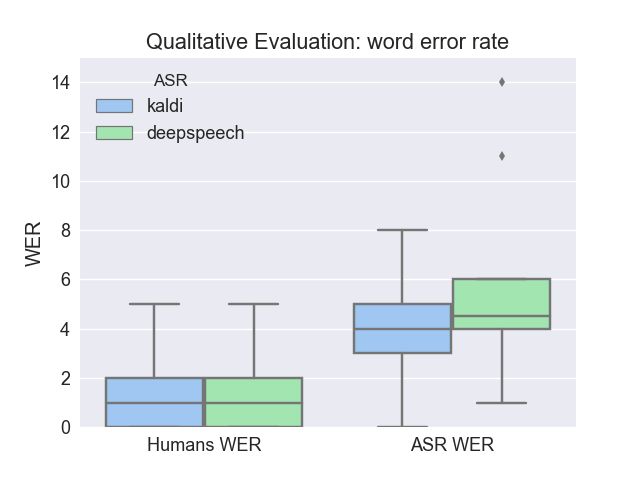}
    \caption{Word Error Rate comparison between human and ASR transcribed text on adversarial samples}
    \label{fig:qual_wer}
\end{figure}

\subsection{Qualitative Evaluation}
While quantitative evaluation empirically proves the goodness of our framework using WER and CC, it is necessary to determine if humans can be fooled with this approach. For this purpose, a qualitative evaluation has been carried out by conducting subjective listening tests with $27$ participants.  

These tests comprised of questions for evaluating an audio sample on 4 aspects: (i) comprehensibility of content, (ii) listening effort, (iii) presence of perturbations, and (iv) naturalness of generated samples. Each participant is presented with $24$ random audio samples (from both targeted and un-targeted attacks) and is required to transcribe the text and rate the listening effort required (from 1-4, with 4 denoting highest effort) to comprehend the text. This provides an indirect measure of the noise/aberrations present in the samples. The WER is calculated for the transcribed text and compared with the ASR generated text, with the original text being the ground truth. We obtained $225$ data points to infer our insights from. Fig. \ref{fig:qual_wer} shows the WER for the two ASR systems and human transcribed text. The mean WER for the human transcribed text is $1.5$ which clearly suggests that humans could not find any significant differences between the adversarial and the original samples. 

\begin{figure}
    \includegraphics[scale=0.45]{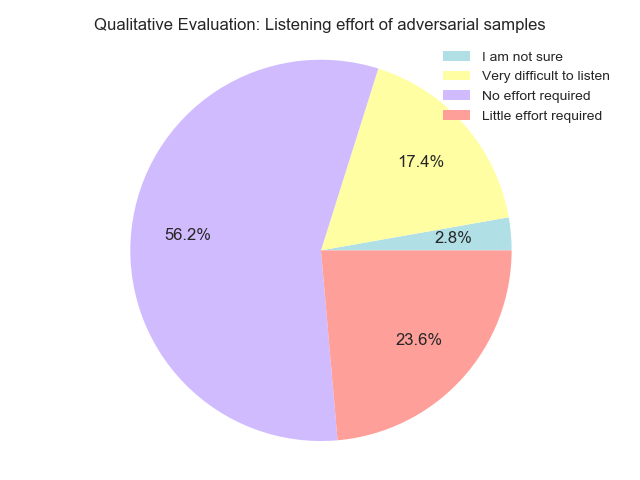}
    \caption{Listening effort required by the survey participants to understand and transcribe the adversarial samples 
    }
    \label{fig:qual_listen}
\end{figure}

Fig. \ref{fig:qual_listen} depicts the listening effort required by the participants to comprehend the audio. More than $80\%$ of the participants required little or no effort to comprehend the text, indicating that there was either no or minimal audible noise in the generated adversarial samples.

To evaluate the naturalness of the generated samples, we asked the participants to distinguish between the original and adversarial samples. They were able to correctly identify both original and adversarial samples 50\% of the time which is equivalent to a random guess. To verify if humans can determine if the generated audio sample has been manipulated, the participants were asked to listen to the generated samples and determine if they were artificially generated or altered in some manner. 73\% said the audio sounded natural, whereas $27\%$ said that they could hear some ticks in the background which made them believe that the audio could be synthesized.

Lastly, we conducted a \textit{transferability} experiment to study if the samples generated to fool one ASR system can do the same to another system. The adversarial samples generated for Deepspeech in an un-targeted setting were tested on Kaldi-ASR and vice versa. In this setting, the mean WER of Kaldi-ASR and Deepspeech are $1.9$ and $3.6$ respectively. This indicates that though there are some common weaknesses between the two ASR systems which can be exploited, there are other model specific weaknesses which may be a result of the model's inductive bias or bias introduced by the training data. This is an interesting research direction to pursue in the future.






%% file: conclusion.tex
\section{Conclusion and Future Work}
In this work, we introduce an algorithm agnostic framework for attacking ASR systems using evolutionary  multi-objective optimization. The framework is tested on two ASR systems and is used for both un-targeted and targeted attacks. Our fitness function decomposes adversarial audio quality evaluation into two objectives: Euclidean distance of MFCC features to get the acoustic similarity between audio samples and edit distance to measure the generated text similarity. We show that MOGA and NSGA-II can be plugged into the framework to generate adversarial samples with high WER and CC. To the best of our knowledge, this is the first time a multi-objective fitness criterion is used to optimize auditory and textual features jointly.